\documentstyle[aps,epsfig]{revtex}
\begin{document}
\title{Chains in critical fluids and nanopores} 

\author{Amina  Negadi$^{1,2}$ Arti Dua$^{1}$ Thomas A. Vigis$^{1}$} 
\address{$^{1}$Max-Planck
  Institut f\"ur Polymerforschung, Ackermannweg 10, D-55122 Mainz, Germany}
\address{$^{2}$ University Aboubakr Belkaid of Tlemcen, Faculty of Sciences,
  Department of chemistry, BP 119, Tlemcen 13000, Algeria}

\date{\today}

\maketitle

\begin{abstract}

\newline
\newline
  The conformational behavior of a polymer in a critical binary solvent
  confined in a porous medium is studied. The size of the polymer in bulk,
  which is mainly governed by the correlation length of the solvent density
  fluctuations, depends on the proximity to the critical point of the binary
  mixture. We find that in contrast to the bulk behavior, the conformational
  properties of the polymer in a porous medium depends strongly on the pore
  size. The latter controls the correlation length of the solvent density
  fluctuations and thus determines the polymer size.
\end{abstract}

\section{Introduction and Model}

The conformational behavior of polymers in a critical (binary) solvent is
mainly governed by two length scales - the chain length and the correlation
length of the solvent density fluctuations \cite{Vilgis:93.1,Brochard:80.1}.
In particular, the size of a polymer in bulk is determined by the proximity to
the critical point of the solvent mixture.  Far away from the critical point,
the size of a polymer in a binary mixture of good solvents simply scales as $R
\sim N^{3/5}$, where $N$ is the number of monomers. Close to the critical
point, however, when the correlation length of the solvent density
fluctuations is comparable to the chain size, the polymer collapses to form a
globule of size $R \sim N^{1/3}$; a collapse transition close to the critical
point is believed to be driven by the solvent density fluctuations which
induce attractive interactions between different parts of the chain. At the
critical point itself, when the solvent density fluctuations act on a much
larger scale compared to the polymer size, the chain regains its original
size, that is, $R \sim N^{3/5}$. These effects first studied by Brochard and
De Gennes \cite{Brochard:80.1} using simple scaling arguments were later
confirmed by simulations of Magda et al.  \cite{Magda:88.1}, field theoretical
and variational studies \cite{Vilgis:93.1,Dua:99.1,Vilgis:98.1} and experiments
\cite{Choi:98.1,Negadi:99.1,Negadi:99.2}.

The fact that the solvent density fluctuations are instrumental in
determining the bulk behavior of polymers in a critical binary solvent 
raises an important question when such a polymer mixture is present in a
restricted geometry, say a porous medium. Since the pore size affects
the correlation length of the solvent density fluctuations and thus
the polymer size, the question as to how a given confinement controls the
conformational behavior of the polymer is an important one. This paper 
addresses this issue, which, we believe, has its implications in  
certain experimental situations like chromatography.

We begin by presenting a brief review of the bulk behavior --- a self-avoiding
polymer chain immersed in a binary solvent mixture, and describe it using the
Edwards Hamiltonian formalism. The idea is to determine the effective
potential exerted on a chain under various conditions of temperature and fluid
composition. The Hamiltonian that describes the system of interest is given by
\begin{eqnarray}
\label{H1}
  \beta {\cal H} & = & \frac{3}{2b^{2}}\int_0^N 
   \left(\frac {\partial{\bf R}(s)}{\partial
    s}\right)^{2} 
ds +\frac{1}{2} v \int_0^N ds \int_0^N ds^{\prime} \delta \left({\bf R}(s)-{\bf
    R}(s^{\prime})\right) \nonumber \\ & + &
\sum_{\sigma=1,2}\sum_{i} v_{\sigma} \int_{0}^{N} ds \delta \left({\bf R}(s)-{\bf
  r}_i^\sigma \right) +  \beta {\cal H}_{\rm f},
\end{eqnarray}
where $\beta = 1/k_{\rm B}T$, $k_B$ is the Boltzmann constant and $T$ the
absolute temperature; ${\bf R}(s)$ is the chain variable, $s$ the curvilinear
coordinates along the chain (contour variable); the chain consists of $N$
segments each of size $b$.  The first term accounts for the connectivity of
the chain and represents the chain entropy due to its elasticity. The second
term results from two body interaction between monomers, where $v$ is the
strength of the bare monomer-monomer excluded volume interaction, which is
assumed to be short ranged. The third term represents the interaction between
the chain and the fluid in which $v_{\sigma}$ is the strength of the short
range excluded volume interaction between the monomers and the different
species of the fluid. $v_{\rm{1}} \neq v_{\rm{2}}$, i.e. the solvent quality
is slightly different. In essence, it means that the chain is likely to be
surrounded by solvent $2$, when $v_{\rm {1}} > v_{\rm{2}}$.
$\beta {\cal H}_{\rm f}$ describes the Hamiltonian of the fluid, the
form of which needs to be specified.  At this point we introduce a
collective density field (order parameter) for the fluid, which is given by
\begin{equation}
\label{den}
c(\bf r) = \sum_{i} \delta({\bf r} - {{\bf r}_{i}}),
\end{equation}
which allows to rewrite the Hamiltonian in the simple form
\begin{eqnarray}
  \beta {\cal H} & = & \frac{3}{2b^{2}}\int_0^N \left(\frac {\partial{\bf R}(s)}{\partial
    s}\right)^{2} 
ds +\frac{1}{2} v \int_0^N ds \int_0^N ds^{\prime} \delta [{\bf R}(s)-{\bf
  R}(s^{\prime})] \nonumber \\ 
& + &(v_{{1}} - v_{{2}})
\sum_{\bf {k}}\int_0^N ds \exp [-i {\bf k} {\bf R}(s)] c({\bf k}) + \beta
{\cal H}_f({c({\bf r})}),
\end{eqnarray}
 where $\beta {\cal H}_{\rm f}({c})$ can now be viewed as
\begin{eqnarray}
\label{Hf}
  \beta {\cal H}_f ({c({\bf r})})  =  \int d^3r \left ( \frac{1}{2} (|
    \nabla c({\bf r}) |^2 + \tau c^2({\bf r})) + \frac{\lambda}{4}
    {c}^{4}({\bf r})\right). 
\end{eqnarray}
The parameter $\tau$ describes the distance from the critical point and
${\lambda}$ is a coupling constant. Far away from the critical point the
quartic term in Eq. (\ref{Hf}) can be neglected, which amounts to $\lambda =
0$. The latter is a Gaussian approximation, which allows us to integrate out
the solvent degrees of freedom to produce an effective interaction potential
between different parts of the chain \cite{Vilgis:93.1}. The effective
Hamiltonian is given by
\begin{equation}
\label{Heff}
\beta {\cal H}_{\rm eff} = \frac{3}{2b^2}\int_0^N \left( \frac{\partial {\bf R}(s)}{\partial s}\right)^{2}ds
+ \sum_{{\bf k}} \int_0^N ds \int_0^N ds^{\prime} \tilde {v}({\bf k}) \exp [- \rm
i{\bf k} ({\bf R}(s)-{\bf R}(s^{\prime}))],
\end{equation}
where  
\begin{eqnarray}
\label{veff} 
\tilde {v}({\bf k})= v -
\frac { (\Delta v)^{2}} {\tau +
(b\bf {k})^{2} } \equiv v - \frac { (\Delta v)^{2}} {(b/\xi_{\rm f})^{2} +
(bk)^{2} } . 
\end{eqnarray}
In writing the above equation, we have defined $\Delta v = v_{1} - v_{2}$.
Since the (mean field) correlation length of the solvent density fluctuations
is defined by $\xi_{\rm f} \simeq b/\sqrt{\tau}$, the form of Eqs.
(\ref{Heff}) and (\ref{veff}) suggests that in certain temperature regimes
depending on the strength of interaction between the chain and the fluid, the
effective potential can become attractive (negative) to induce a chain
collapse. In other words, depending on the correlation length of the solvent
density fluctuations and $\Delta v$, the effective solvent quality can become
poor. Therefore, before we discuss the problem of a chain in a restricted
geometry, it is useful to briefly review the behavior of a chain in a nanopore
under poor solvent conditions.

\section{Nanopores}
\subsection{Chains in Nanopores}
Let us first review some of the scaling results of chains in nanopores. The
diameter $D$ of the tubes is assumed to be smaller than the natural chain
radius $R_{\rm g} \simeq N^{\nu}$, where in three dimensions
$\nu = 1/2$ for theta-solvents and $\nu \approx 3/5$ for good solvents.

We consider flexible polymer chains confined in pores of cylindrical shape
with neutral walls such that there is no adsorption between the polymer and
the pores; the volume available to the solvent and the solute is limited by
 well defined boundaries. For the case of a single chain in a capillary of
diameter $ D$, there are two distinct behaviors depending on the dimensionless
ratio $R_{F}/D$, where $R_{F}$ is the Flory radius of a polymer in bulk.  If
$R_{F} < D$, we have a conventional bulk solution. If $R_{F} > D$ the chain is
deformed and is confined in a cylindrical tube of diameter $D \ll R_{F}$.
Since $D \gg b$, the chain still retains some lateral wiggling. We assume that
the tube wall repel the chain strongly so that there is no trend towards
adsorption.  The length of the tube occupied by the chain is $R_{\parallel}$.
The behavior of the chain can mainly be understood with the blob picture. The
size of the blob is solely determined by the tube diameter.  The parallel size
of the chain is then made out of $n_{\rm b}$ blobs of diameter $D$ such that
$R_{\parallel} = n_{\rm b} D$. The only information about the quality of the
solvent is then in the number of blobs themselves, which is given by $n_{\rm
  b} = N/g$, where $g$ is the number of monomers inside a blob. The latter is
simply given by $g = (D/b)^{1/\nu}$, and the size of the chain along the tube
is $R_{\parallel} \simeq b N (b/D)^{(1-\nu)/\nu}$.  Note that the same results
can be derived by using a free energy as $\beta{\cal {F}} =
{R_{\parallel}^2}/{N b^2} + \beta U$, where $\beta U \simeq b^{3} {N^2}/{D^2
  R_{\parallel}}$ for good solvents and $\beta U \simeq b^{6}{N^3}/{(D^2
  R_{\parallel})^{2}}$ for theta-solvents.

As discussed in the last section, the solvent density fluctuations can 
induce effective attractive interaction between different parts of the 
chain segments leading to a chain collapse. Since this amounts to the
effective solvent quality becoming poor, it is worthwhile to consider
the case of a polymer in a poor solvent confined in a nanopore. The free
energy of such a system is given by
\begin{eqnarray}
\label{Fps}
\beta{\cal {F}} \simeq \frac{R_{\parallel}^{2}}{Nb^{2}} -
|v| \frac{N^2} {D^2 R_{\parallel}} + w
\frac{N^3}{\left (D^2 R_{\parallel} \right)^2},
\end{eqnarray}
where $v\sim - t b^3$ and $ w \sim b^6$ are the two and three body
interactions respectively. Here the reduced temperature $t$ describes the
distance from the theta temperature, i.e., $t = |T-\Theta |/ \Theta$. Let us
first discuss the influence of the tube geometry on the $\Theta$ - point. To
estimate the $\Theta-$ temperature,
we employ the usual Ginzburg criterion. The $\Theta-$point for a
finite chain length in a restricted geometry cannot be defined by
simply setting the second virial coefficient equal to zero. Instead, a
consistent expression can be determined by the balance between the chain entropy and the
attractive second virial term. In particular, when $t$ is sufficiently
large (the solvent quality is poor enough), the first term in Eq.
(\ref{Fps}), which represents the entropic elasticity, is smaller than
the attractive interaction due to the second term. Thus the
$\Theta-$ point for chains inside the pore can be determined by comparison
between the entropic term and the negative second virial term. 
This yields the following condition for the $\Theta$- point:
\begin{equation}
\label{teta0}
t \simeq (D/b)^{2} (R_{\parallel}/bN)^{3},
\end{equation}
where the estimate is sensitive to the unperturbed chain size
$R_{\parallel}$. The usual Ginzburg criterion uses the ideal chain size $R
\propto bN^{1/2}$, but this yields the unphysical result $ t_{\rm c} \simeq
(1 / \sqrt N ) ( {D}/b N^{1/2} )^{2} $ since it suggests an increase
in $t$ with the increase in the pore diameter. Moreover, the chain
length cannot play a significant role since the collapse in the pore takes place only on the length scales less than the pore
diameter, i.e., the relevant blob size. Therefore, it is  physical to use
the Gaussian value for $R_{\parallel}$ inside the pore, which is
simply given by $R \simeq bN (b/D)$, as suggested by a simple blob
argument. As a result, the value for the shift of the $\Theta$ - temperature is given by
\begin{equation}
\label{teta}
t = \frac{b}{D}.
\end{equation}
The above equation shows significant dependence of the
$\Theta$-temperature on the pore size. It can be rewritten as
\begin{equation}
\theta = \theta_{\rm bulk} + {\rm const} \; {b\over D},
\end{equation}
which is only valid for $D \leq b\sqrt N$, i.e., when the pore diameter is
smaller compared to the Gaussian chain size. When $D$ becomes comparable to
the chain size, then the shift of the $\Theta$ - temperature is determined by
$t \simeq b/R_{\rm Gauss}$, which is simply $t \propto 1/\sqrt N$, as it must
be the case for the collapse of  polymer chains in bulk.
 The geometry dependence of the $\Theta$-temperature
has recently been studied for a polymer confined between two parallel
plates (slit geometry) by using numerical methods \cite{mishra}. For
completeness, we would like to remark that naive scaling for a simple slit
geometry predicts $t_{\rm c, slit} = (b/D)^{0}$, which suggests a logarithmic
dependence, $t_{\rm c, slit} \propto \log (D/b)$, but this needs further
investigation. 

For $t > t_{\rm c}$ the chain becomes significantly
contracted from its theta size. The stretching term can then be neglected, and
the balance between the attractive and repulsive third body potential yields
the chain size
\begin{eqnarray}
R_{\parallel} \sim \frac{b}{t} N \left ( \frac{b}{D} \right )^2.
\end{eqnarray}
The density inside the tube is simply given by $\rho \sim \frac{N}{D^2
  R_{\parallel}} \simeq t /b^{3}$, which corresponds to the density of a
unconstrained globule and also defines the thermal blobsize $\xi_{T} = b/t$.
On length scales smaller than $\xi_{T}$, the chain statistics are unperturbed
by the volume interactions and are that of a random walk, i.e., $\xi_{T} \sim
b g_{T}^{1/2}$. The parallel extension of the chain can, therefore, be
expressed by comparing the blobsize $\xi_{T}$ and the diameter $D$,
\begin{equation}
R_{\parallel} \sim b N \frac{\xi_T}{D}  \left ( \frac{b}{D} \right ).
\end{equation}
Using the above expression, the size of the polymer confined in a tube of
diameter $D$ can be summarized for three distinct cases: When $D < \xi_T$, the
chain experiences theta conditions; when $D > \xi_T$, the chain inside the
tube finds itself in a poor solvent condition. It collapses to form a globule
of size governed by $t$. The chain starts to shrink when $D = \xi_{T}$,
starting from the size $R_{\parallel} \sim b N t$, i.e., a linear
arrangement of the thermal blobs.

\subsection{Critical fluids in Nanopores}

The problem of phase transitions in restricted geometry has been considered
mostly in the case of (two dimensional) wetting. In a nanopore the situation
is very different. The problem of a spherical model in a spherocylinder has
been studied by Cardy \cite{cardy}, who shows using conformal mapping that
the correlation length at the critical point can be written as
\begin{equation}
\xi^{-2} = A/D^{2}
\end{equation}
where $A = d-2+\eta$ is a universal amplitude; it depends only on the space
dimension $d$ and the critical exponent $\eta$. this results implies also that
the amplitude and so $\xi$ does not depend directly on the value of the fluid
coupling constant $\lambda$, which  was introduced in eq. (\ref{Hf}).
Physically, this implies that
the criticality plays a significant role only on the scale of the tube
diameter $D$ and fluctuations always stay finite. Therefore, the effects of
the solvent density fluctuations on the conformational behavior of a polymer
chain, which is dissolved in a bicomponent solvent present in a porous medium,
are mainly local. These effects are studied in the next section.

\subsection{Polymer chain and fluids in nanopores}

As discussed in the previous section, when the chain dissolved in a
bicomponent solvent is confined in a nanopore, the problem is effectively one
dimensional. To see this we can rewrite all spatial vectors in their
Cartesian coordinates, that is
\begin{eqnarray}
  \beta {\cal H} & = & \frac{3}{2b^{2}}\sum_{\mu = x,y,z}\int_0^N \left(\frac {\partial{R_{\mu}}(s)}{\partial
    s}\right)^{2} 
ds +\frac{1}{2} v  \int_0^N ds \int_0^N ds^{\prime} \prod_{\mu = x,y,z} \delta [R_{\mu}(s)-
  R_{\mu}(s^{\prime})] \nonumber \\ 
& + &(v_{{1}} - v_{{2}})
\sum_{\bf {k}}\int_0^N ds \exp [-i \sum_{\mu = x,y,z} k_{\mu}R_{\mu}(s)] c({\bf k}) + \beta
{\cal H}_f({c({\bf r})}).
\end{eqnarray}
 Moreover, when the constraints due to the cylindrical geometry of
the tube are imposed the effective potential remains
structurally similar.  Because of the confinement, the vector ${\bf k}$ can be
splitted into its Cartesian components, i.e., ${{\bf k}} = (k_{\parallel},
k_{\perp}, k_{\perp})$, where $k_{\parallel}$ corresponds to the length scale
along the (cylindrical) tube. The component $k_{\perp}$ is limited by the tube
wall and can be replaced by its upper limit $1/D$. This naive estimate
suggests that the effective potential is given by
\begin{eqnarray} 
\tilde{v}({\bf k})= v -
\frac { (v_{ 1} - v_{ 2} )^{2}} {(bk_{\parallel})^{2}+\tau+\left
    (\frac{b}{D} \right)^2}. 
\end{eqnarray}
Within this mean field model, the new correlation length can the be identified
as
\begin{equation}
b^{2}\xi^{-2}_{\rm tube} = \tau + \left(\frac{b}{D} \right)^{2}.
\end{equation}
It is interesting to note that the scaling of the correlation
length agrees with the exact result of Cardy \cite{cardy} at $\tau = 0$.

To estimate the size of the chain within the classical scaling, it is
sufficient to discuss the limit $k_{\parallel} = 0$. The effective
potential can then be casted into a Flory form, which yields together
with the Gaussian stretching term a reasonable estimate for the size
of the chain. The effective potential is given by
\begin{eqnarray} 
\tilde{v}= v -
\frac{(\Delta v )^{2}} {\tau + \left (\frac{b}{D}\right )^2}.
\end{eqnarray}
For $\tau = 0$, i.e., at the critical point this equation is similar to the
corresponding mean-field equation in the bulk, Eq. (\ref{veff}) in the low
wave vecotr limit. For geometrically unrestricted systems $D \to \infty$, the
correlation length grows then with the pore diameter, because here we have
$\xi_{\rm f} \propto D$. At the critical point, i.e., $\tau = 0$, the effective
monomer-monomer potential $\tilde{v}$ becomes negative when the tube diameter
is sufficiently large:
\begin{eqnarray} 
\left({D\over b} \right)^{2} \geq \frac{v}{(v_{1}-v_{2})}.
\end{eqnarray}
Since this amounts to the effective interaction between the chain segments
being attractive, the chain is expected to collapse inside the tube. In
contrast to the bulk behavior, where in a certain temperature regime (close to
the critical point) the chain is always collapsed, the collapse in a confined
geometry depends purely on material properties --- the strength of the
polymer - fluid interaction, the chain excluded volume, and the diameter of the
tube. The geometry, therefore, plays an important role in the collapse of a
polymer chain both in a critical fluid and a fluid with a finite correlation
length.

\subsection{The size of the chain}

The Hamiltonian as given by Eq. (\ref{Heff}) can be transformed into a Flory
free energy by simple dimensional analysis {\cite{Vilgis:00.1}}. It has to be
modified such that the $3$-dimensional Dirac function is anisotropic and the
lateral dimensions are determined by the diameter $D$ of the nanopore.  Thus,
we can estimate the excluded volume to be $\delta ({\bf R}(s)-{\bf R}(s'))
\propto 1 / (D^2 R_{\parallel})$, where $R_{\parallel}$ is the chain size
parallel to the pore. The interaction parameter between the monomers is
determined by the value for the effective chain potential $ \tilde
v(k_{\parallel} = 0)$:
\begin{eqnarray}
\beta {\cal F} = \frac{R_{\parallel}^2}{N b^2} + \tilde {v}(k_{\parallel} = 0)
\frac{N^2} {D^2 R_{\parallel}}
\end{eqnarray}
Minimization of the above free energy yields the size of the chain in the
nanopore:
\begin{eqnarray}
R_{\parallel} \sim \left ( v -\frac { (v_{1} - v_{2} )^{2}}
  {\tau +\left (\frac{b}{D}\right )^2
  }\right)^{1/3} N \left ({\frac{b}{D}}\right)^{2/3}. 
\end{eqnarray}
The above expression is valid as long as the effective potential is positive.
At the critical point $\tau = 0$, we can use Cardy result for $d=3$
and to get
\begin{equation}
R_{\parallel} \simeq b N 
\left(1 - {1\over {(1+\eta)}}\chi^{2} \left( \frac{D}{b}\right)^{2}  \right)^{1/3}
\left({b\over D} \right)^{2/3},
\end{equation}
where we have rescaled the interaction parameters by the size of the Kuhn
length $v \sim b^{3}$ and introduced $\chi^{2} = (v_{1} - v_{2})^{2}$. This
results holds as long as the effective potential is positive, which means that
the pore size is small enough. For larger pore sizes as
\begin{equation}
D > \frac{v}{(v_{1}-v_{2})^{2}},
\end{equation}
the effective potential becomes negative; the critical fluid then resembles a
poor solvent. The size of the chain can then be estimated by taking into
account the three body interaction in the tube. It lateral extension is then
given by 
\begin{equation}
R_{\parallel} \propto N {1 \over {(\chi^{2} (D/b)^{2} - 1)}}
\left(\frac{b}{D}\right)^{2}.
\end{equation}
Therefore, the chain is always in a stretched conformation, as has
already been seen from the scaling analysis of the behavior of chains
in solvents of different quality.

\section{Applications and Discussion}

The behavior of chains immersed in a critical fluid confined in a
porous media is substantially different from the bulk case. As
mentioned in the Introduction, a chain in bulk undergoes a
conformational transition, i.e., $N^{3/5} \to N^{1/3} \to N^{3/5}$, on
approach to the critical point of the fluid. In a restricted geometry, 
however, the correlation length of the fluid stays finite and is
mainly determined by the typical size of the tube. The latter implies
that the pore diameter can select chains according to their chain
length --- for a given pore diameter $D$, short chains (of size much
less than $D$) are not restricted by the confinement; they can undergo
a collapse transition on approach to the critical point, as is the case in the
bulk. Sufficiently long chains (of size much larger than $D$), on the
contrary, remain in a stretched conformation. As a result, the variation in the
fluid temperature, that is, approach towards its ``criticality'' can
select chain conformations according to their polymerization degree. 

To see this let us consider a dilute solution of chemically identical
chains having degrees of polymerization $N_{1}$ and $N_{2}$ dissolved
in a good solvent such that $N_{1}>>N_{2}$.
In terms of the present model, a good solvent quality implies $\tau >
\chi^{2} - (b/D)^{2}$ (i.e. the temperature not too close to the critical
point where $\tau = 0$). When the pore size is such that $D \sim
{N_{2}^{3/5}}$, only long chains of size ${N_{1}^{3/5}}$ will be
confined. The confinement free energy for the long chain is given by
\begin{equation}
\beta F \simeq N_{1} \left (\frac{b}{D}\right)^{5/3},
\end{equation}
which is estimated as the number of blobs times $k_{B}T$. The free
energy for the non-confined chain is roughly given by
\begin{equation}
\label{nc}
\beta F \approx N_{2}\log N_{2} - (\gamma -1) \log N_{2}.
\end{equation}
where $\gamma$ is the partition function critical exponent. The
comparison of the two free energies yields
\begin{equation}
{D\over b} \approx \left ( {N_{1}\over N_{2}}\right)^{3/5},
\end{equation}
which provides a simple scaling for the separation of chains
according to their chain length. It is to be noted that in writing the 
above expression we have ignored the logarithmic dependence in Eq. (\ref{nc}).

Eq.(\ref{teta}) suggests  another interesting possibility --- the chain
length dependence of the second virial coefficient allows us to
determine the temperature away from the critical point where the
separation between the collapsed short chains and the more stretched long
chains occur:  
\begin{equation}
\tau^{*} = \frac{\chi^{2}}{1+(b/ D)} - 
\left( {b\over D}\right)^{2}.
\end{equation}
For chains whose size is less than that of the pore diameter, the
above condition suggests
\begin{equation}
\tau_{N}^{*} = \frac{\chi^{2}}{1+1/\sqrt N}
\end{equation} 
Thus by tuning the distance from the critical point, that is $\tau$, the
chains may adopt different (globular) conformation depending on their chain
lengths. This provides a means for separating chains by changing the poor
solvent quality (through the change in the criticality of the fluid) in porous media.

\section*{Acknowledgment}
A. N. acknowledges the financial support of the German Academic Exchange
Service DAAD for her stay in Mainz. We thank K. Kremer for useful discussions.


\end{document}